# A pedestrian level strategy to minimize outdoor sunlight exposure in hot summer

Xiaojiang Li [a, b], Yuji Yoshimura [a,c], Wei Tu [a, d], Carlo Ratti [a]

[a] *Senseable City Lab, Department of Urban Studies and Planning, Massachusetts Institute of Technology, Room 9-250, 77 Massachusetts Avenue, Cambridge, MA, 02139, United States*
[b] *Department of Geography and Urban Studies, Temple University, Philadelphia, PA, 19122, United States*
[c] *Research Center for Advanced Science and Technology, The University of Tokyo, 4-6-1 Komaba, Meguro-ku, Tokyo 153-8904, Japan*
[d] *Shenzhen Key Laboratory of Spatial Information Smart Sensing and Services and Research Institute of Smart Cities, School of Architecture and Urban Planning, Shenzhen University, Shenzhen 518060, China*

Email: lixiaojiang.gis@gmail.com

**Abstract:** Too much sunlight exposure would cause heat stress for people during the hot summer, although a minimum amount of sunlight is required for humans. Unprotected exposure to Ultraviolet (UV) radiation in the sunlight is one of the major risk factors for skin cancer. Mitigating the heat stress and UV exposure caused by too much sunlight exposure becomes a pressing issue in the context of increasing temperature in urban areas. In this study, we proposed an individualized and short-term effective strategy to reduce sunlight exposure for urban residents. We developed a routing algorithm minimizing pedestrian's outdoor sunlight exposure based on the spatio-temporal distribution of sunlight in street canyons, which was generated by the simulation of sunlight reaching the ground using Google Street View panoramas. The deep convolutional neural network-based image segmentation algorithm PSPNet was used to segment the GSV panoramas into categories of sky, trees, buildings, and road, etc. Based on the GSV image segmentation results, we further estimated the spatio-temporal distribution of sunlight in street canyons by projecting the sun path over time on the segmented GSV panoramas. The simulation results in Shibuya, Tokyo show that the routing algorithm can help to reduce human sunlight exposure significantly compared with the shortest path. The proposed method is highly scalable and can be easily extended to other cities with GSV data available. This study would provide a pedestrian level strategy to reduce the negative effects of sunlight exposure for urban residents in hot summers.

Keywords: Sunlight exposure; Shade routing; Google Street View (GSV).

## 1. Introduction

Too much sunlight exposure is thermally uncomfortable and potentially dangerous for people during hot summer (Brash et al., 1991; Hodder and Parsons, 2007, Kurazumi et al., 2013, Richards and Edwards, 2017; Young, 2009), although the minimum amount of sunlight is required for humans. Sunlight exposure is a major factor of human heat stress, which is a comprehensive parameter influenced by humidity and temperature (Gaffen and Ross, 1998; NOAA, 2009). Heat stress in summer poses a daily health threat to urban residents and causes morbidity and mortality (Stone et al., 2010). The temperature increases and the increasingly prominent urban heat island effects increase the threat of heat stress on public health. Other than heat stress, unprotected exposure to Ultraviolet (UV) radiation in the sunlight is one or the major risk factors for skin cancers (Brash et al., 1991; Armstrong and Kricker, 2001). Preventing too much sunlight exposure would help people to decrease the potential dangers caused by sunlight.

Urban streets that carry most of the human outdoor activities in cities (Li et al., 2017) are the major place for human outdoor sunlight exposure. Understanding the spatio-temporal distribution of sunlight in street canyons would help us to develop methods to protect people from too much outdoor sunlight exposure. The solar radiation reaching the street canyons is influenced by the obstruction effects of street trees and buildings (Carrasco-Hernandez et al., 2015; Hwang et al.,

2011; Johansson and Erik, 2006; Lin et al., 2010). Buildings and trees on both sides of streets would obstruct sunlight and then reduce the potential human sunlight exposure in street canyons (Li et al., 2018). The urban form and the orientation of the street canyons would influence the obstruction effects of buildings and tree canopies on the sunlight (Algeciras et al., 2016; Ali-Toudert and Mayer, 2006; Johansson, 2006; Zhao et al., 2008). Usually, those streets with larger height-weight ratio have less incoming sunlight reaching the ground. Those east-west orientation streets have more solar radiation reaching the ground because of the same direction with the sun's zenith (Sanusi et al, 2016). In different time of one day, the different sun positions and the surrounding obstructions would make sunlight exposure significantly different.

The high-resolution building height models make it possible to simulate the obstruction effects of building blocks on sunlight and estimate the transmission of sunlight within street canyons, which further make it possible to generate the spatio-temporal distributions of sunlight in the street canyons. However, building height models usually oversimplify the geometries of urban canyons (Carrasco-Hernandez et al., 2015). In addition, building height models only consider the shading effects of buildings and the shading effects of tree canopies and other urban features are usually not considered (Li et al., 2018a).

The street-level images provide a new approach to simulate the sunlight within street canyons. Different from the building height models, which usually only include the information about building blocks, the street-level images can represent all types of obstructions along the streets (Li et al., 2018a). Therefore, using street-level images would be more reasonable to simulate the shading effects of obstructions and generate more accurate spatio-temporal distributions of sunlight in street canyons. In addition, street-level images, such as, Google Street View images and Mapillary images, are globally available and publicly accessible. All of these make street-level images suitable for predicting human outdoor sunlight exposure along streets.

In this study, we used the Google Street View (GSV) images as a surrogate of the streetscape environment to simulate the sunlight in street canyons. A pre-trained deep learning model was used to segment the street-level images and recognize the obstructions of sunlight within street canyons. We then generated the spatio-temporal distribution of sunlight in street canyons by calculating the sun positions over times and projecting sun positions on the segmented GSV images. Based on the generated spatio-temporal distribution of sunlight exposure in the study area, we further developed a spatio-temporal routing algorithm to provide an individualized routing choice for people to minimize their sunlight exposure.

## 2. Study area and dataset

The study area, Shibuya, is a special ward and a major commercial and business center in Tokyo, Japan. Shibuya is very densely populated with an estimated population of 221,800 and population density of 14,679.09 people per $km^2$. In order to collect GSV images in the study area, we created sample sites along streets every 10m (**Fig. 1**(a)). The street map used in this study was collected from Open Street Map. The coordinates of these sample sites were then used to collect the metadata (**Fig. 1**(b)). This study focuses on human sunlight exposure in hot summer, therefore, only those images captured in similar seasons were used in this study. Based on the time stamps in the collected metadata of GSV panorama, we only selected the most recently captured images in summer for each sample site and downloaded GSV panoramas (**Fig. 1**(c)).

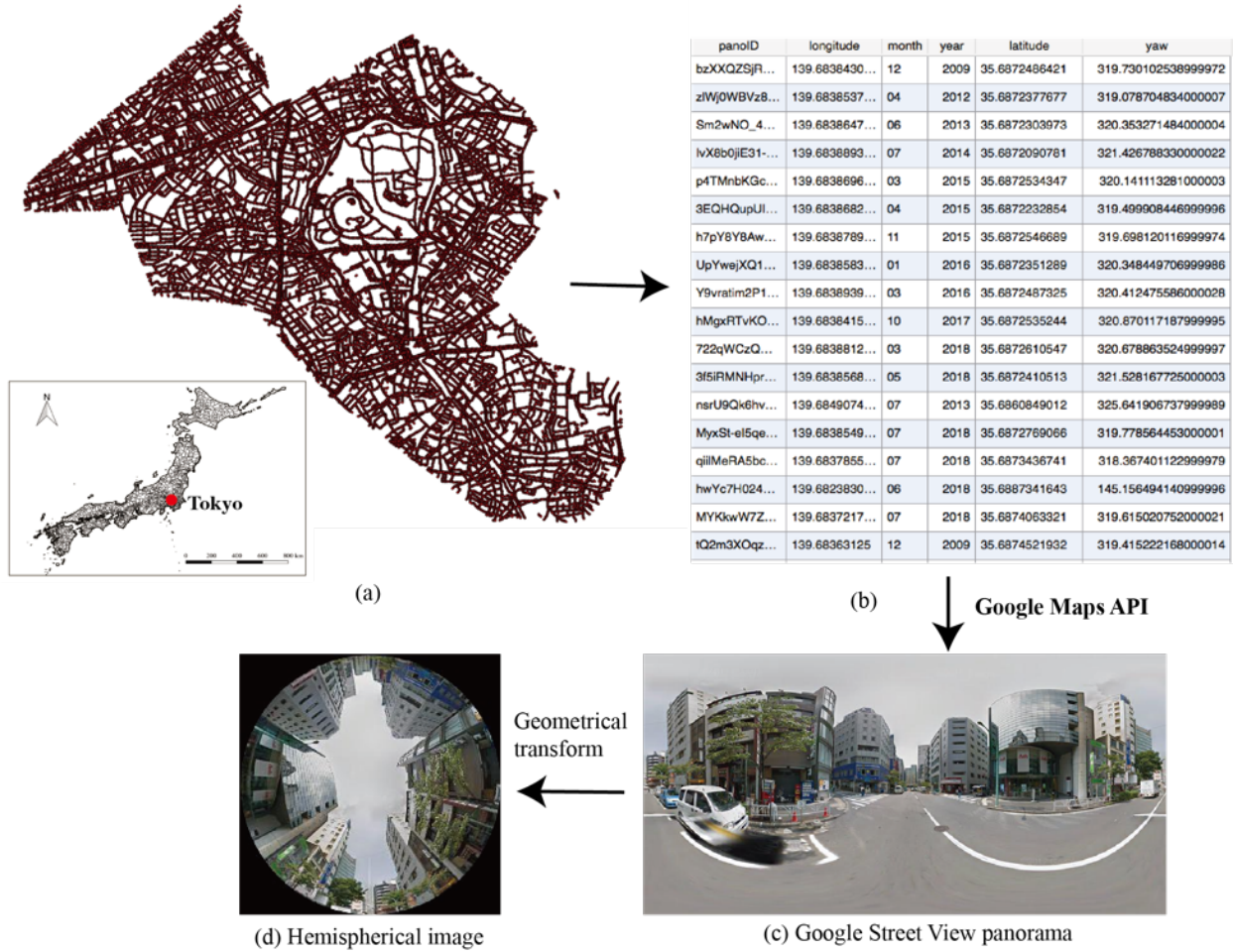

**Fig. 1.** The workflow for Google Street View (GSV) panoramas collection in Shibuya, Tokyo, Japan, (a) the created sample sites in the study area, (b) the metadata of GSV panoramas, (c) a downloaded GSV panorama, (d) a generated hemispherical image based on geometrical transform.

## 3. Methodology

3.1 Hemispherical images generation and segmentation

Hemispherical image-based method is one of the standard methods for estimating human sunlight exposure on the ground (Li et al., 2018a, b). In this study, hemispherical images were generated from cylindrical GSV panoramas by geometrical transform (Li et al., 2018a; Li et al., 2017). In order to derive the sunlight obstruction information in street canyons, the image segmentation algorithm PSPNet was used to segment GSV panoramas into sky pixels and obstruction pixels (Vegetation, buildings, and impervious surface pixels) (Gong et al., 2018; Zhao et al., 2016). The segmented GSV panoramas were further geometrically transformed into hemispherical images, which were used to model the human sunlight exposure within the street canyons. **Fig. 2** shows the segmentation results (**Fig. 2**(b)) of PSPNet on three GSV panoramas (**Fig. 2**(a)) and the segmented hemispherical images based on GSV panorama segmentation results (**Fig. 2**(c)).

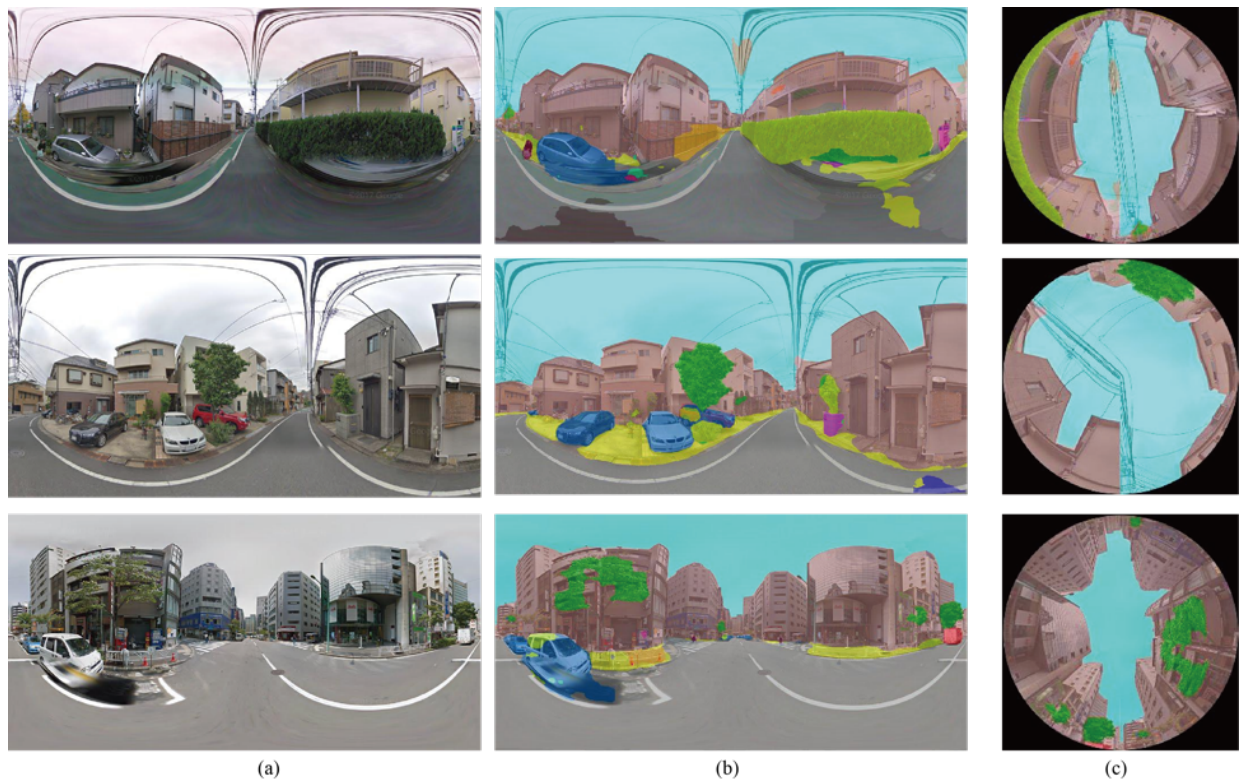

**Fig 2.** Image segmentation results using PSPNet, (a) original Google Street View (GSV) panoramas, (b) the blend of segmentation results on GSV panoramas, (c) the hemispheric view of the segmented GSV panoramas.

3.2 Human exposure to sunlight in street canyon

A pedestrian's exposure to sunlight would be influenced by the time, orientation of the streets, buildings, and street tree canopies. Based on the generated segmented hemispherical images, it is possible to estimate whether a pedestrian is exposed to sunlight or not at any time and location by projecting the sun position on the hemispherical images.

In this study, we implemented the sun position estimation algorithm developed by NOAA Earth System Research Laboratory (ESRL, https://www.esrl.noaa.gov/) to estimate the sun position at any specific time. **Fig. 3** (a) depicts the geometrical model of the sun and a pedestrian. **Fig. 3**(b) shows the projected sun positions in one day on three hemispherical images in the study area. For a person at the location of (*lon*, *lat*), if the sun at one time (*t*) is projected on sky pixels of the hemispherical images, then the person is exposed to direct sunlight at that time. If the sun position is on non-sky pixels, the person at that time is shaded from sunlight.

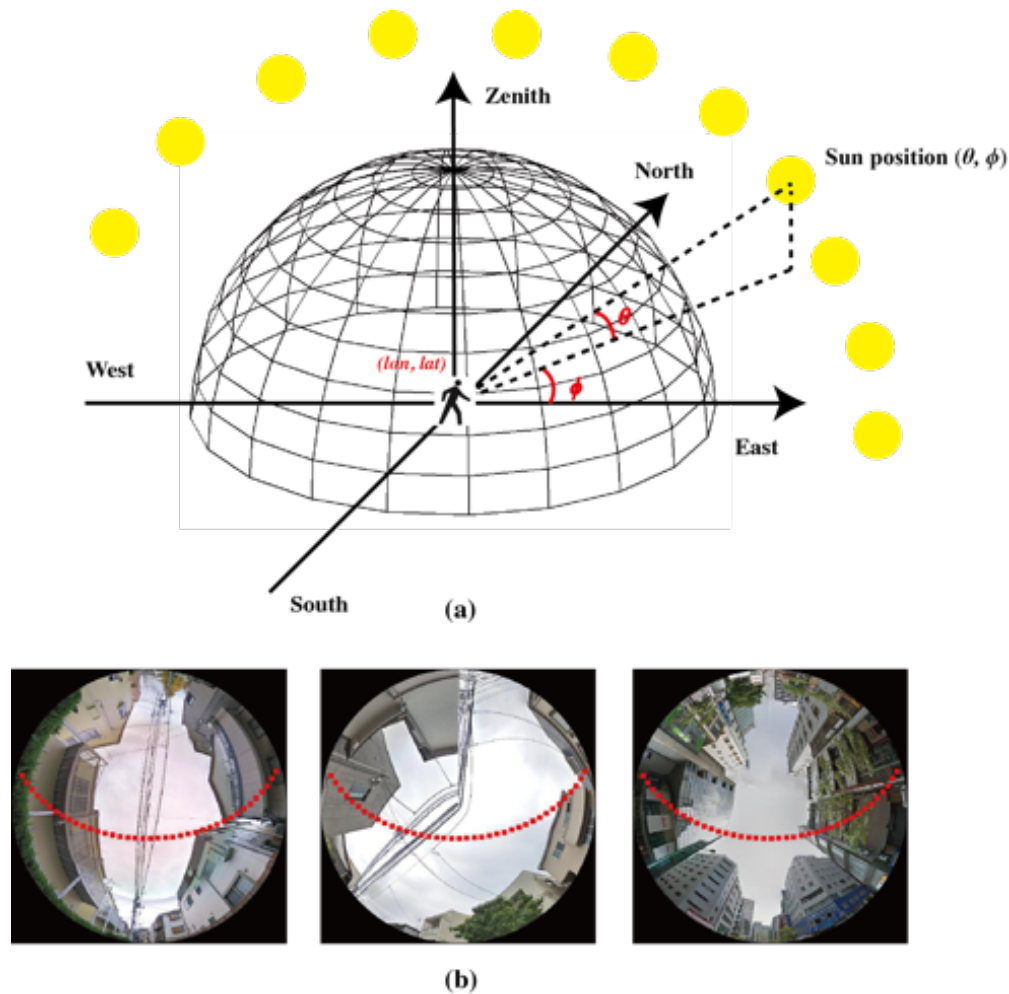

**Fig. 3.** The geometrical model of the sun (a) and the overlay of sun path in one day on hemispherical images (b).

The intensity of the sunlight at different times of one day varies with the sun elevation angle. In this study, we calculated the weighted sunlight exposure ($E_{w,t}$) at time $t$ as,

$$E_{w,t} = B_t \cos \theta_t \qquad (1)$$

Where $\theta_t$ is the sun elevation angle at time $t$; $B_t$ is the Boolean variable indicating whether the direct sunlight is obstructed or not at time $t$; if the sunlight is obstructed at time $t$, the $B_t$ equals 0, or the $B_t$ is 1. The variations of solar radiation due to cloudiness and other atmospheric conditions are not considered in this study.

3.3 Routing algorithm for minimizing sunlight exposure

Based on the weighted sunlight exposure, a person's accumulated sunlight exposure $E_a$ from one location at time $t_0$ to another location at time $t_n$ can be estimated as,

$$E_a = \sum_{t=t_0}^{t_n} E_{w,t} = \sum_{t=t_0}^{t_n} B_t \cos \theta_t \qquad (2)$$

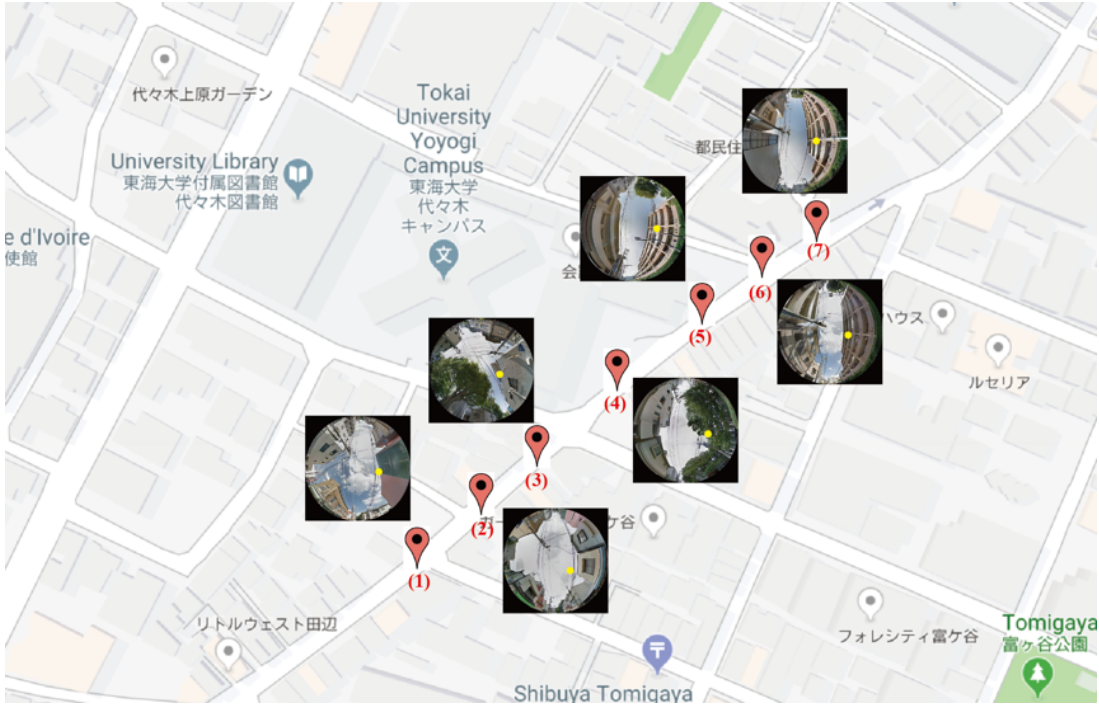

**Fig. 4.** Human exposure to sunlight along one route.

**Fig. 4** shows a sequence of hemispherical images with corresponding sun positions overlaid along a street of the study area.

The "exposure over distance" parameter *α*, which indicates the trade-off between sunlight exposure and distance was used to trade-off between distance and sunlight exposure. Therefore, the routing algorithm will find the minimum accumulated sunlight exposure (Min $E_a$) from several route candidates.

$$Minimize\ E_a = \sum_{t=t_0}^{t_n} dist \cdot [\alpha B_t cos\theta_t + (1-\alpha)] \qquad (3)$$

Where *dist* is the distance between two nearby stop points along the route. The trade-off parameter *α* as 0 indicates the shortest geographical distance path from origin to destination. The trade-off parameter *α* as 1 indicates route with the minimum sunlight exposure. In this study, we set the trade-off parameter *α* as 0.5.

Routing for the minimum exposure to sunlight is a time-dependent routing issue, in which the weight of the graph change over time. Therefore, we first generated the sunlight exposure distribution along streets every 5 minutes. We assume the sunlight exposure at each street segment is not change that much in 5 minutes. We then applied Dijkstra (1959) algorithm with considering the temporal change of the sunlight exposure for different street segments in the study area.

## 4. Results

Based on the metadata of all available GSV panoramas, we only selected the most recent images captured in leaf-on seasons (April, May, June, July, August, September, and October) in this study. Finally, we collected 45,085 GSV panoramas along streets in the study area. **Fig. 5** shows the spatial distribution of the finally collected GSV panoramas and the time stamps of those downloaded GSV panoramas. Generally, the GSV panoramas cover most streets in the study area. There is a small region in the northern part of the study area, which is the Meiji Shrine, has no GSV images coverage. Considering the relatively isolated location of the shrine, we think the unavailability of GSV images in the small portion of the study area will have no much influence on the results of the routing algorithm in the study area.

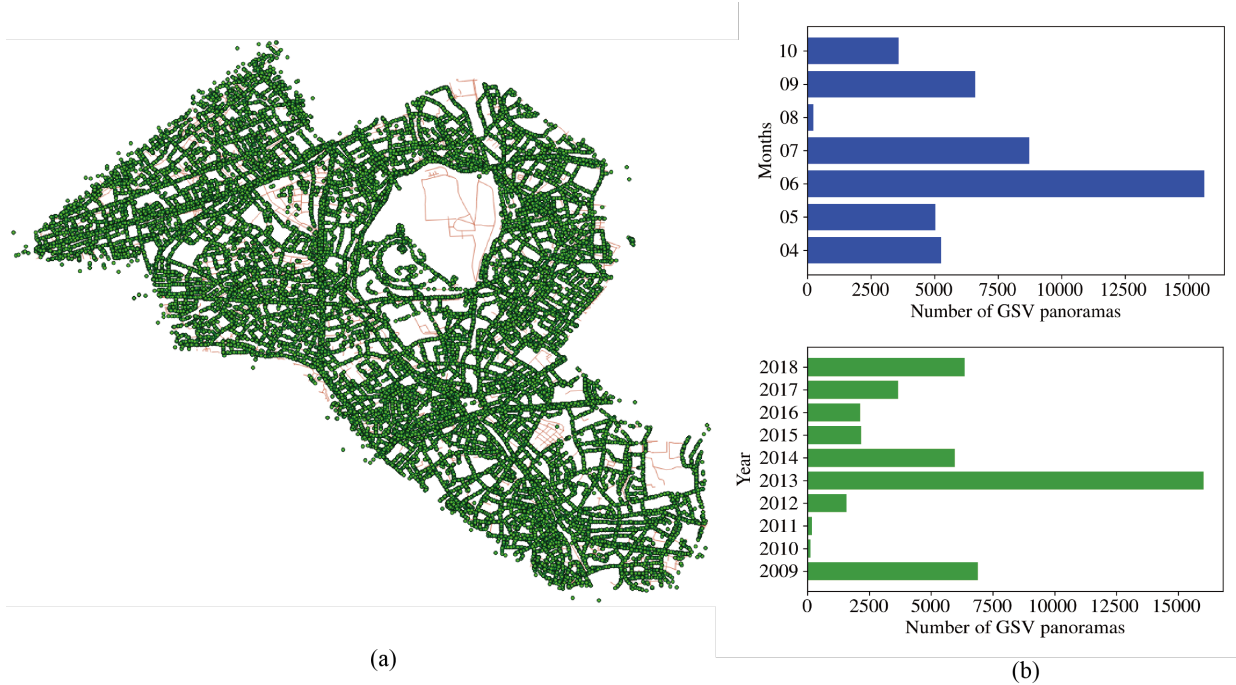

(a)  (b)

**Fig. 5.** The spatial distribution of the downloaded Google Street View (GSV) panoramas (a) and time stamps (b) of those downloaded GSV panoramas.

Based on the segmented hemispherical images and predicted sun positions over time in the study area, we generated the spatio-temporal distribution of the sunlight exposure at the point level for every 5 minutes. We then aggregated the point level sunlight exposure map to street segments by assigning each point to the closest street segments. The sunlight exposure attribute for each street segment is the mean value of all assigned point. **Fig. 6** shows the spatial distribution of street-level sunlight exposure over time on July 15$^{th}$, 2018 at 9:00 am, 12:00 pm, 2:00 pm, and 5:00 pm. It can be seen clearly that at 9:00 am and 5:00 pm, those East-West orient streets are exposed to sunlight directly, and streets with other orientations are not directly exposed to sunlight. This is because at sunrise morning and sunset afternoon the sun zenith matches the East-West streets.

9:00 am

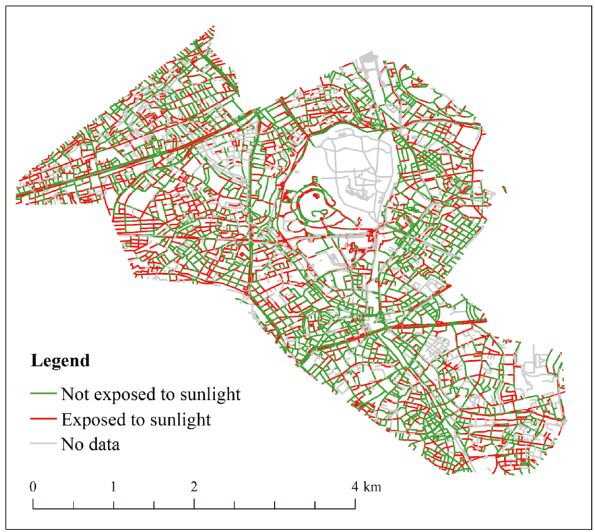

12:00 pm

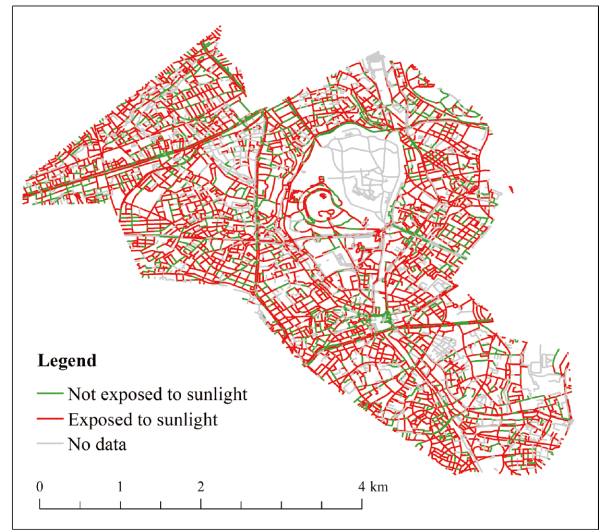

2:00 pm

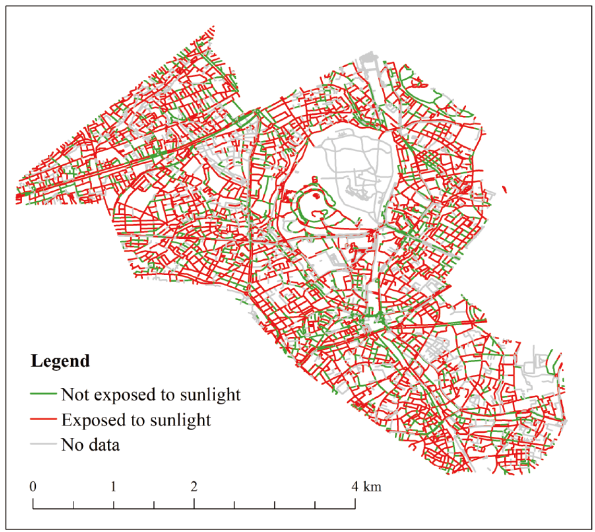

5:00 pm

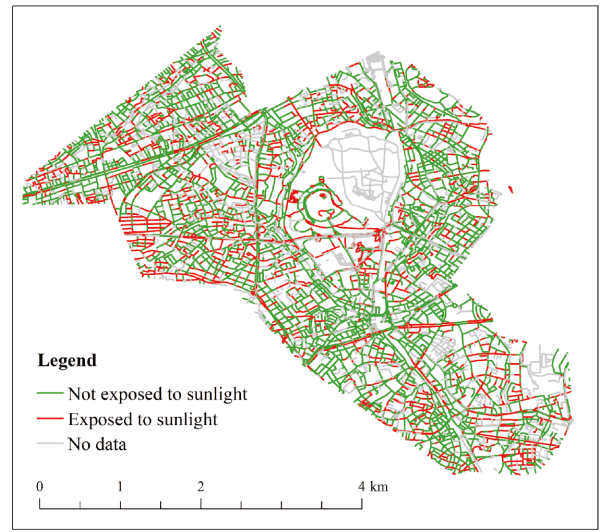

**Fig. 6.** The spatial distributions of street-level Boolean sunlight exposure variable at 9:00 am, 12:00 pm, 2:00 pm, and 5:00 pm on July 15$^{th}$, 2018 of the study area.

In order to compare the results of the shortest geographical distance path and the minimum sunlight exposure path, we randomly created 1,000 origin-destination pairs with random starting time from 9 am to 6 pm. Results show that the minimum sunlight exposure paths help to decrease the potential sunlight exposure by 35.23% compared with the shortest geographic distance path.

**Fig. 7** shows four of these 1,000 paths with the minimum sunlight exposure and trade-off parameter as 0.5 and corresponding paths with the shortest geographic distance from origins to destinations at the starting time of 11 am on July 15$^{th}$, 2018.

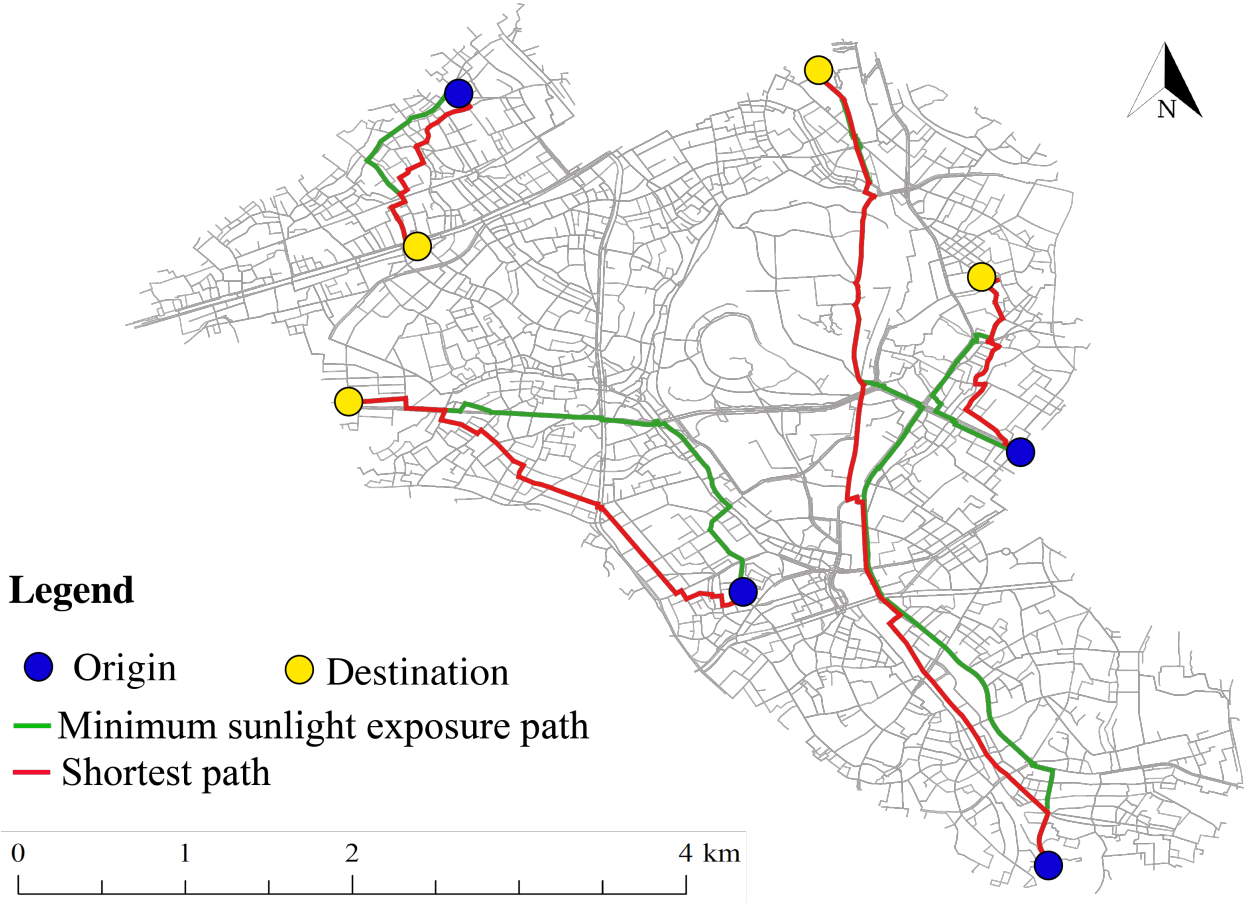

**Fig. 7.** The paths of minimum sunlight exposure (green line) and the path of shortest geographical distance (red line) of four origin-destination pairs at the randomly starting time from 9 am to 6 pm in July 15$^{th}$, 2018.

## 5. Discussion

Too much sunlight exposure would cause human heat stress and unprotected sunlight exposure could cause skin cancers, which is one of the most common cancers. Understanding the spatio-temporal distribution of sunlight exposure would help us to develop strategies to increase thermal comfort and reduce environmental hazards from the sunlight in cities. There are many methods have been developed and applied to protect human beings from too much sunlight exposure, such as planting trees to increase shade, wearing sunscreen, installing shelters, etc. Different from those plans, this study proposed a pedestrian level strategy to protect people from too much outdoor sunlight exposure during hot summer. The Google Street View (GSV) panoramas and deep convolution neural network were used to map the spatio-temporal distribution of the sunlight exposure with consideration of the sun positions and obstruction effects of buildings and tree canopies.

By analyzing the sunlight exposure of 1,000 pairs of randomly created origins and destinations in the study area, results show that the proposed routing algorithm can help to reduce the potential sunlight exposure by 35.23% on July 15$^{th}$, 2018 compared with the shortest geographical distance path. Although the exact number of sunlight exposure reduction would vary for different days and different daily routes of people, this study shows the possibility to significantly reduce the sunlight exposure using the proposed method. This study provides a new way at the individual level to deal with the risks of too much sunlight exposure. The proposed method would also help to democratize the environmental information to fight with the negative effects of hazards from sunlight to urban dwellers based on the publicly accessible GSV images. The proposed method provides a bottom-up solution to deal with heat stress and unprotected UV exposure for people. The proposed method would let individuals have the information on their daily sunlight exposure. With such information available, individuals can change their daily schedule or choose the best route to protect themselves from too much sunlight exposure. This study would also provide an important reference for urban planners and city governments to reduce human sunlight exposure during hot summers by street designs, such as providing shades, increase street tree canopies, etc.

The proposed method can generate more accurate sunlight exposure information compared with other digital building height model-based methods since street-level images were used to represent the streetscapes. Using the street-level image-based method is more reasonable to consider the actual human sunlight exposure in street canyons because all kinds of sunlight obstructions are considered using the street-level image based method. In addition, the proposed method is highly scalable for estimating the sunlight exposure over time considering the publicly accessible and globally available of street-level images.

This study still has some limitations on modeling human sunlight exposure. The street-level image-based method may not be able to perfectly represent pedestrian's sunlight exposure since GSV images were collected in the central part of streets, not sidewalks. Considering the fact that the streets in the study area are very narrow, therefore, it is quite reasonable to use the street-level images to represent pedestrian's exposure to the sunlight. However, the method should be adjusted for other study areas with wide streets and separate sidewalks.

In addition, human thermal stress is not only influenced by the sunlight exposure but also the humidity, wind speed, air temperature, etc. In this study, only the direct sunlight exposure is considered. The sunlight exposure in street canyons varies spatially and temporally much more than humidity and wind speed. For one specific site if we assume the wind speed and humidity are constant, therefore, the exposure to the sunlight would be considered as the most important contributing factor of the human thermal comfort. Therefore, it is reasonable to use the sunlight exposure as a surrogate to develop routing algorithm to maximize the thermal comfort for a pedestrian walking from one place to another place.

This study used the randomly created sample sites to understand the performance of the proposed routing algorithm for reducing the sunlight exposure. In order to better understand the performance of the routing algorithm on local people, future study may also need to consider the human daily GPS trajectories in order to better understand how much sunlight exposure can be reduced.

Although the routing algorithm in this study provides an individualized strategy to mitigate the urban heat island and global climate change in cities, in long term, cutting carbon emission, increasing green space, decrease impervious surfaces should be considered to deal with those urban environmental issues.

In future study, the human thermal comfort during winter will also be considered since sunlight would be considered as comfortable in winter.

# 6. Conclusions

This study proposed a bottom up approach for understanding the spatio-temporal distribution of sunlight exposure and a pedestrian level strategy for people to mitigate outdoor heat stress during hot summers. The Google Street View panoramas and a deep learning-based image segmentation algorithm were used to derive streetscape environment and model the human sunlight exposure within street canyons. Considering the global availability of street-level images around the world, the proposed method would provide us an effective method in dealing with the heat stress in the context of global temperatures rising and urban heat island. This study would also provide urban planners with important information about human thermal comfort and inform them of taking effective measures to increase the thermal comfort in cities.